\documentclass[11pt,a4paper]{article}

\usepackage{bm,amsfonts}
\usepackage{graphicx}
\usepackage{hyperref}

\newcommand{\Z}{\mathbb{Z}}
\newcommand{\C}{\mathbb{C}}
\newcommand{\R}{\mathbb{R}}
\newcommand{\PP}{\mathbb{P}}
\newcommand{\N}{\mathcal{N}}

\newcommand{\U}{\mathrm{U}}
\newcommand{\SU}{\mathrm{SU}}
\newcommand{\SO}{\mathrm{SO}}
\newcommand{\GL}{\mathrm{GL}}
\newcommand{\Hom}{\mathrm{Hom}}
\newcommand{\diag}{\mathrm{diag}}
\newcommand{\s}{\bar{s}}

\makeatletter

\@addtoreset{equation}{section}
\makeatother

\date{\today}

\begin{document}

\begin{titlepage}

\renewcommand{\thefootnote}{\fnsymbol{footnote}}

\begin{flushright}
RIKEN-MP-28
\\
\end{flushright}

\vskip5em

\begin{center}
 {\Large {\bf 
 $\beta$-ensembles for toric orbifold partition function
 }}

 \vskip3em

 {\sc Taro Kimura}\footnote{E-mail address: 
 \href{mailto:kimura@dice.c.u-tokyo.ac.jp}
 {\tt kimura@dice.c.u-tokyo.ac.jp}}

 \vskip2em

{\it Department of Basic Science, University of Tokyo, 
 Tokyo 153-8902, Japan\\ \vskip.2em
 and\\ \vskip.2em
 Mathematical Physics Lab., RIKEN Nishina Center, Saitama 351-0198,
 Japan 
}

 \vskip3em

\end{center}

 \vskip2em

\begin{abstract}
We investigate combinatorics of the instanton
 partition function for the generic four dimensional toric orbifolds.
It is shown that the orbifold projection can be implemented by taking
 the inhomogeneous root of unity limit of the $q$-deformed partition function.
The asymptotics of the combinatorial partition function yields the
multi-matrix model for a generic $\beta$. 
\end{abstract}

\end{titlepage}

\tableofcontents

\setcounter{footnote}{0}


\section{Introduction}
\label{sec:Intro}

The instanton counting is extensively applied to various
non-perturbative aspects of the four dimensional gauge theory.
In particular Nekrasov partition function
\cite{Nekrasov:2002qd,Nekrasov:2003rj} plays an essential role not
only in the four dimensional Seiberg-Witten theory
\cite{Seiberg:1994rs,Seiberg:1994aj}, but also the two dimensional
conformal field theory.
The remarkable connection between the four and two dimensional theories
through the instanton partition function
is called AGT relation \cite{Alday:2009aq}, and generalized to various
situations, for example, the higher rank theory
\cite{Wyllard:2009hg,Mironov:2009by}, the asymptotic free theory
\cite{Gaiotto:2009ma,Marshakov:2009gn,Taki:2009zd} and also the ALE
space
\cite{Belavin:2011pp,Nishioka:2011jk,Bonelli:2011jx,Belavin:2011tb,Bonelli:2011kv},
etc.

The ALE space is given by resolving the singularity of the orbifold
$\C^2/\Gamma$ \cite{Eguchi:1978xp,Gibbons:1979zt,Kronheimer:1989zs}.
The instanton construction
\cite{springerlink:10.1007/BF01233429,springerlink:10.1007/BF01444534},
the instanton counting \cite{Fucito:2004ry} and the wall-crossing
\cite{Nishinaka:2011nn,Nishinaka:2011is} are considered in this case as
well as the Euclidean space $\R^4 \simeq \C^2$.
Furthermore the inhomogeneous orbifold
theory is discussed in terms of the AGT relation \cite{Kanno:2011fw}:
it is shown that the instanton counting on the inhomogeneous orbifold
$\C \times \C/\Z_r$ is utilized to describe the theory in the presence of a
generic surface operator.
Not only the four dimensional theory, but also the two dimensional
theory with vortices on orbifolds has been recently investigated
\cite{Kimura:2011wh}.

In this paper we develop the previous result \cite{Kimura:2011zf}, and
consider a systematic method to deal with the combinatorial
representation of the partition function for the generic four
dimensional toric orbifolds $\C^2/\Gamma_{r,s}$, whose boundary is the
generic lens space $L(r,s)$.
It includes the type $A_{r-1}$ ALE space as $\C^2/\Gamma_{r,r-1}=\C^2/\Z_r$.
So far $\N=4$ theories on such a space, and also Chern-Simons theory on
the lens space $L(r,s)$ have been investigated
\cite{Fucito:2006kn,Griguolo:2006kp,Brini:2008ik,Gang:2009wy}.
Recently further research is done with respect to the index, and its
relation to the three and two dimensional theories \cite{Benini:2011nc}.

We can obtain the orbifold partition function by performing the orbifold
projection for the standard one.
However it is apparently written in a complicated form, 
we will show a much simpler method to assign the orbifold projection.
To implement that we first lift it up to the
$q$-deformed theory, and then take the root of unity limit of it, as well as
the standard orbifold $\C^2/\Z_r$ discussed in Ref.~\cite{Kimura:2011zf}.
The similar method is also applied to the spin Calogero-Sutherland model
\cite{Uglov:1997ia} (see also Ref.~\cite{KuramotoKato200908}).

We also discuss the $\beta$-ensemble matrix model for the
toric orbifold theories.
We successfully obtain the multi-matrix model with a generic $\beta$
by extracting the asymptotics of the combinatorial partition
function \cite{Klemm:2008yu,Sulkowski:2009br,Sulkowski:2009ne}.
The $\Omega$-background parameter is related to this parameter as
$\beta=-\epsilon_2/\epsilon_1$, so that it is important to discuss the
generic $\beta$-ensemble in order to consider the application to the AGT
relation.
When we estimate the asymptotic behavior of the combinatorial part
corresponding to the matrix measure, we need the root of unity limit of the
$q$-deformed Vandermonde determinant, which is the weight function for
the Macdonald polynomial \cite{Mac_book}.
This suggests that we can obtain a new kind of polynomials induced from
the Macdonald polynomial by taking this limit.

This paper is organized as follows.
In section \ref{sec:count} we consider the ADHM construction for the
toric orbifolds $\C^2/\Gamma_{r,s}$, and then obtain the
combinatorial expression for the partition function.
We will show that the root of unity limit of the $q$-deformed partition
function is essential for the orbifold projection, and it is useful to
introduce the basis of the fractional exclusive statistics.
Section \ref{sec:matrix} is devoted to derivation of the matrix models.
We obtain the $\beta$-ensemble multi-matrix model by taking the
asymptotic limit of the combinatorial partition function.
In section \ref{sec:summary} we summarize the results with some
discussions.

\section{Instanton counting on toric orbifolds}\label{sec:count}

Let us start with the generic four dimensional toric space.
It is given by the quotient $\C^2/\Gamma_{r,s}$ where $\Gamma_{r,s}$ is
a $\Z_{r}$ action labeled by the two coprime integers $(r,s)$ with
$0<s<r$ as
\begin{equation}
 \Gamma_{r,s}:~ (z_1, z_2) 
  \quad \longrightarrow \quad
  ( \omega_r z_1, \omega_r^s z_2)
  \label{orb_action1}
\end{equation}
where $\omega_r = \exp (2\pi i/r)$ is the primitive $r$-th root of
unity.
This space goes to the lens space $L(r,s)$ at infinity.

The orbifold action (\ref{orb_action1}) generates a singularity at the
origin of $\C^2$.
We can obtain the smooth manifold by resolving the singularity, which is
called the Hirzebruch-Jung space \cite{BarthPetersVandeven}.
After blowing up the singularity there are $\ell$ two-spheres
characterized by the generalized Cartan matrix
\begin{equation}
 C = \left(
      \begin{array}{ccccc}
       - e_1 & 1 & 0 & \cdots & 0 \\
       1 & - e_2 & 1 & \cdots & 0 \\
       0 & 1 & - e_3 & \cdots & 0 \\
       \vdots & \vdots & \vdots & \ddots & \vdots \\
       0 & 0 & 0 & \cdots & - e_\ell \\
      \end{array}
     \right)
 \label{Cartan}
\end{equation}
where the self-intersection numbers $e_i$, $i=2, \cdots, \ell$ are obtained
by expanding the rational number $r/s$ in a continued fraction form
\begin{equation}
 \frac rs =
  e_1-{1\over\displaystyle e_{2}- {\strut
  1\over \displaystyle e_{3}- {\strut 1\over\displaystyle\ddots {}~
  e_{\ell-1}-{\strut 1\over e_\ell}}}}
\end{equation}
and $e_1$ is the smallest integer greater than $r/s$.
In the case of the ALE space, namely $s=r-1$, we have $e_i=2$ and
$\ell=r-1$.
Therefore the matrix (\ref{Cartan}) coincides with the Cartan matrix for the
type-$A_{r-1}$ Lie algebra.

We then consider the standard ADHM construction for $\R^4 \simeq
\C^2$ to study instanton counting before orbifolding.
The ADHM equations for $k$-instanton configuration for $\SU(n)$ theory
are given by
\begin{eqnarray}
 \mathcal{E}_{\C} & := & 
  \left[
   B_1, B_2
  \right] + IJ = 0, \\
 \mathcal{E}_{\R} & := &
  [ B_1, B_1^\dag ] + [ B_2, B_2^\dag ]
  + I I^\dag - J^\dag J = 0
\end{eqnarray}
where the ADHM data $(B_1, B_2, I, J)$ are interpreted as elements of
homomorphisms,
\begin{equation}
 B_1, B_2 \in \Hom \left(V,V\right), \qquad
 I \in \Hom \left(W,V\right), \qquad
 J \in \Hom \left(V,W\right).
\end{equation}
The rank of the gauge group and the instanton number are encoded in
dimensions of the vector spaces, dim~$V=n$ and dim~$W=k$, respectively.
Actually, when we consider $\SU(n)$ theory, 
we had better deal with $\U(n)$ group, and then implement the condition for
the Coulomb moduli $\sum_{l=1}^n a_l=0$.
Note that this procedure is not enough for some cases: we have to factor
out the $\U(1)$ contribution when we consider the AGT relation \cite{Alday:2009aq}.

There is $\U(k)$ gauge symmetry for these ADHM data
\begin{equation}
 \left(B_1, B_2, I, J \right) \quad \longrightarrow \quad
 \left( g B_1 g^{-1}, g B_2 g^{-1}, g I, J g^{-1}  \right), \qquad
 g \in \U(k),
\end{equation}
and thus the instanton moduli space is given by
\begin{equation}
 \mathcal{M}_{n,k} = \left\{(B_1, B_2, I, J)| 
		      \mathcal{E}_{\C}=0, \mathcal{E}_{\R}=0 \right\} / \U(k).
 \label{mod_sp1}
\end{equation}
The resolution of singularity of this ADHM moduli space is given by the
following quotient \cite{Nakajima:2003uh},
\begin{equation}
 \widetilde{\mathcal{M}}_{n,k} = \left\{(B_1, B_2, I, J)| 
	   \mathcal{E}_{\C}=0, \mbox{stability cond.} \right\} // \GL(k,\C).
 \label{mod_sp2}
\end{equation}
The stability condition is interepreted as the irreducibility for the
moduli space.

\begin{figure}[t]
 \begin{center}
  \includegraphics[width=35em]{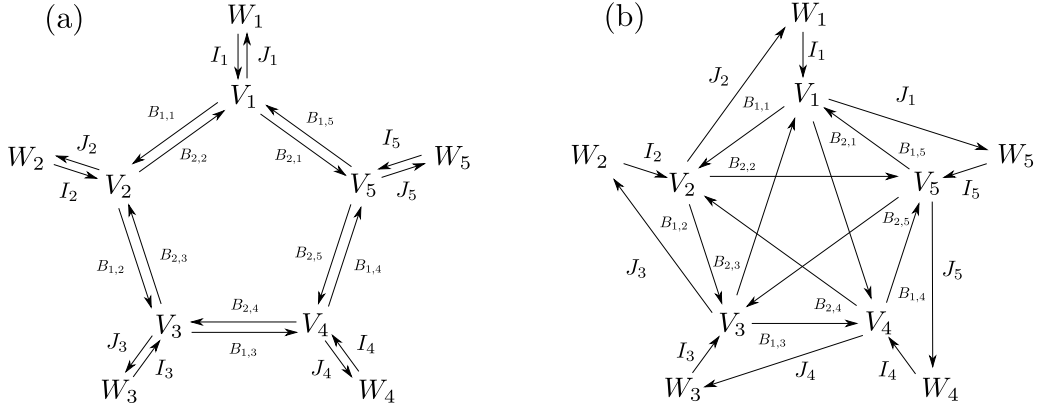}
 \end{center}
 \caption{Quiver diagrams for orbifolding ADHM data:
 (a) $\C^2/\Gamma_{5,4}$ ( the ALE space $\C^2/\Z_5$) and (b)
 $\C^2/\Gamma_{5,3}$.}
 \label{fig_quiv}
\end{figure}

We then consider the action of isometries on $\C^2$ for the ADHM data
\begin{equation}
  \left(B_1, B_2, I, J \right) \quad \longrightarrow \quad
 \left( T_{1} B_1 , T_2 B_2, I T_{a}^{-1}, T_1 T_2 T_a J  \right)
\end{equation}
where $T_a = \diag(e^{ia_1}, \cdots, e^{ia_n}) \in \U(1)^{n}$,
$T_{\alpha} = e^{i \epsilon_\alpha} \in \U(1)^2$.
They are the torus actions coming from the symmetry of $\U(n)$ and
$\SO(4)$, respectively.
We have to consider the fixed point of these isometries up to gauge
transformation $g\in\U(k)$ to perform the localization formula.
Thus the orbifold action, corresponding to (\ref{orb_action1}), on the
ADHM data is
\begin{equation}
 \Gamma_{r,s}:~ (B_1, B_2, I, J) \quad \longrightarrow \quad
  (\omega_r B_1, \omega_r^s B_2, I, \omega_r^{1+s} J).
  \label{orb_action2}
\end{equation}
Due to the orbifold action we have to
introduce decomposed vector spaces with respect to the irreducible
representations of $\Z_r$,
\begin{equation}
 W = \bigoplus_{v=1}^{r} W_v, \qquad
 V = \bigoplus_{v=1}^{r} V_v,
\end{equation}
Here we assign the orbifold action for the gauge group element
as $e^{ia_l} \to \omega_r^{p_l}e^{ia_l}$.
It is just a holonomy, which characterizes the
boundary condition of the gauge field.
Thus the ADHM data surviving under the orbifold action can be
written as
\begin{equation}
 B_{1,v} \in \Hom(V_v, V_{v+1}), \quad
 B_{2,v} \in \Hom(V_v, V_{v+s}), \quad
 I_v \in \Hom(W_v, V_v), \quad
 J_v \in \Hom(V_v, W_{v+1+s}),
\end{equation}
and thus we have
\begin{eqnarray}
 \mathcal{E}_{\C} & \longrightarrow &
  B_{1,v+s} B_{2,v} - B_{2,v+1} B_{1,v} + I_{v+1+s} J_v, 
  \label{momentum_map1} \\
 \mathcal{E}_{\R} & \longrightarrow &
  B_{1,v-1} B_{1,v-1}^\dag - B_{1,v}^\dag B_{1,v}
 + B_{2,v-s} B_{2,v-s}^\dag - B_{2,v}^\dag B_{2,v}
 + I_v I_v^\dag - J_v J_v^\dag .
 \label{momentum_map2}
\end{eqnarray}
They are periodic modulo $r$ as $W_{r+1}=W_1$ and so on.
Fig.~\ref{fig_quiv} shows quiver diagrams for orbifolding ADHM data.
These conditions are much complicated, and thus the whole
structure of the instanton moduli space is not yet clear.
Actually (\ref{momentum_map1}) and (\ref{momentum_map2}) take non-zero
value after resolving the singularity.
It can concern possibility of the localization method.
Its availability is investigated for the case of the ALE space
\cite{Fucito:2004ry} and more generic theories
\cite{Gasparim:2008ri,Bruzzo:2008}, but we have to consider this problem
more explicitly.
Basically we still have the isometry $\U(1)^2$ corresponding to the
spatial rotation even for the orbifolds.
This might ensure that we can apply the localization formula to these cases.

We then derive the combinatorial representation of the partition function.
Since the characters of the vector spaces are given by
\begin{equation}
 V = \sum_{l=1}^n \sum_{(i,j)\in\lambda^{(l)}} T_{a_l} T_1^{1-i} T_2^{1-j},
  \qquad
 W = \sum_{l=1}^n T_{a_l},
\end{equation}
the character of the tangent space at the fixed point under the
isometries, which is labeled by $n$-tuple partition $\vec{\lambda}$,
turns out to be
\begin{eqnarray}
 \chi_{\vec{\lambda}} & = & - V^* V (1-T_1) (1-T_2) + W^* V + V^* W T_1 T_2 
  \nonumber \\ 
 & = & \sum_{l,m}^n \sum_{(i,j)\in\lambda^{(l)}}
  \left(
   T_{a_{ml}} T_1^{\lambda_i^{(m)}-j+1} T_2^{-\check\lambda_j^{(l)}+i}
 + T_{a_{lm}} T_1^{-\lambda_i^{(m)}+j} T_2^{\check\lambda_j^{(l)}-i+1}
  \right).
\end{eqnarray}
Here we define $a_{lm} = a_l-a_m$, $T_{a_{lm}}=e^{ia_{lm}}$, etc.
We can extract the weight, and thus obtain the partition function in
a combinatorial way,
\begin{equation}
 Z_{\vec{\lambda}}
  = 
  \prod_{l,m}^n \prod_{(i,j)\in\lambda^{(l)}}
  \frac{1}{a_{ml} + \epsilon_1 (\lambda_i^{(m)}-j+1) - \epsilon_2
  (\check\lambda_j^{(l)}-i)}
  \frac{1}{a_{lm} - \epsilon_1 (\lambda_i^{(m)}-j) + \epsilon_2
  (\check\lambda_j^{(l)}-i+1)}.
\end{equation}
This is the instanton partition function \cite{Nekrasov:2002qd}.
The two parameters, $\epsilon_1$ and $\epsilon_2$, are called
$\Omega$-background parameters, which are required for regularizing the
singularities in the moduli space.

We then have the partition function for the orbifold theory by 
taking into account only the invariant sector under the orbifold action
(\ref{orb_action2}).
Since each contribution to the character behaves under the orbifold action as
\begin{equation}
 \Gamma_{r,s}:~
 T_{a_{ml}} T_1^{\lambda_i^{(m)}-j+1} T_2^{-\check{\lambda}_j^{(l)}+i}
 \quad \longrightarrow \quad
 \omega_r^{p_{ml}+\lambda_i^{(m)}-j+\s(\check{\lambda}_j^{(l)}-i)+1}
 T_{a_{ml}} T_1^{\lambda_i^{(m)}-j+1} T_2^{-\check{\lambda}_j^{(l)}+i}
\end{equation}
and so on, the $\Gamma_{r,s}$-invariant sector is different for the first
and the second parts in the product,
\begin{equation}
  \frac{1}{a_{ml} + \epsilon_1 (\lambda_i^{(m)}-j+1) - \epsilon_2
  (\check\lambda_j^{(l)}-i)}
  \quad \mbox{for} \quad
  p_l - p_m + \lambda_i^{(m)} - j + \s (\check\lambda_j^{(l)}-i) + 1
  \equiv 0~(\mbox{mod}~r),
  \label{gamma_inv1}
\end{equation}
\begin{equation}
  \frac{1}{a_{lm} - \epsilon_1 (\lambda_i^{(m)}-j) + \epsilon_2
  (\check\lambda_j^{(l)}-i+1)}
  \quad \mbox{for} \quad
  p_{l} - p_m + \lambda_i^{(m)} - j + \s (\check\lambda_j^{(l)}-i) + \s
  \equiv 0~(\mbox{mod}~r),
  \label{gamma_inv2}
\end{equation}
with $\s = r - s$.
Thus the partition function for $\U(1)$ theory on the toric orbifold is
given by
\begin{equation}
 Z_{\lambda;\Gamma_{r,s}} = 
  \prod_{\mbox{\scriptsize $\check\Gamma$-inv.}\subset\lambda}
  \frac{1}{\lambda_i - j + \beta(\check\lambda_j - i) + 1}
  \prod_{\mbox{\scriptsize $\hat\Gamma$-inv.}\subset\lambda}
  \frac{1}{\lambda_i - j + \beta(\check\lambda_j - i) + \beta}.
  \label{partfunc_U1_orb}
\end{equation}
We introduce another parameter defined as $\beta=-\epsilon_2/\epsilon_1$.
Here $\check\Gamma$- and $\hat\Gamma$-invariant sectors stand for the
conditions shown in (\ref{gamma_inv1}) and (\ref{gamma_inv2}), respectively.
It can be easily extended to $\SU(n)$ gauge theory.


These conditions to extract the $\Gamma_{r,s}$-invariant sectors are apparently
complicated, but there is a simple way to implement the orbifold
projection \cite{Kimura:2011zf,Uglov:1997ia}.
To implement the orbifold projection, let us start with the standard
partition function before orbifolding, and then lift it to
the $q$-deformed partition function, which is interpreted as the five
dimensional function,
\begin{equation}
 Z_\lambda^q = \prod_{(i,j)\in\lambda} 
  \frac{1}{1-q^{\lambda_i-j+1}t^{\check\lambda_j-i}}
  \frac{1}{1-q^{-\lambda_i+j}t^{-\check\lambda_j+i-1}}.
\end{equation}
These $q$ and $t$ are related to the $\Omega$-background parameters as
$q=e^{\epsilon_1}$, $t=e^{-\epsilon_2}=q^\beta$.
Of course we obtain the original four dimensional function by taking the
usual $q \to 1$ limit.
On the other hand, by taking the root of unity limit of the $q$-deformed
function, the orbifolded partition function (\ref{partfunc_U1_orb}) is
automatically obtained up to constants.
In this case we assign the following parametrization,
\begin{equation}
 q \longrightarrow \omega_r q, \qquad
 t \longrightarrow \omega_r^{-s} q^\beta = \omega_r^{\bar{s}} q^\beta
 \label{qt_parametrize}
\end{equation}
and then take the limit $q \to 1$. 
To regularize the singular behavior at $q \to 1$, we now take into
account the adjoint matter contribution whose mass parameter is given by
$\mathfrak{m}$.
Thus the weight function yields, for example, 
\begin{eqnarray}
 && \frac{1 - \omega_r^{\lambda_i-j+\bar{s}(\check{\lambda}_j-i)+1} 
  q^{\lambda_i-j+1+\mathfrak{m}}t^{\check{\lambda}_j-i}}
  {1 - \omega_r^{\lambda_i-j+\bar{s}(\check{\lambda}_j-i)+1} 
  q^{\lambda_i-j+1}t^{\check{\lambda}_j-i}}
  \nonumber \\
 & \longrightarrow &
 \left\{
  \begin{array}{ccc}
   \frac{\lambda_i-j+\beta(\check{\lambda}_j-i)+1+\mathfrak{m}} 
        {\lambda_i-j+\beta(\check{\lambda}_j-i)+1} 
    & \mbox{if} & \lambda_i-j+\bar{s}(\check{\lambda}_j-i)+1 \equiv 0~
    ( \mbox{mod}~ r) \\
    1 & \mbox{if} & \lambda_i-j+\bar{s}(\check{\lambda}_j-i)+1 \not\equiv 0~
    ( \mbox{mod}~ r) \\
  \end{array}
 \right. . 
\end{eqnarray}
If we want to extract only the contribution of the vector multiplet, we
have to take the decoupling limit $\mathfrak{m} \to \infty$.

The $q$-partition function for $\SU(n)$ theory can be written with the
cut off parameter $N^{(l)}$ for the number of entries of the partitions
as follows,
\begin{equation}
 Z_{\vec\lambda}^q = \prod_{(l,i)\not=(m,j)}
  \frac{(Q_{lm}q^{\lambda_i^{(l)}-\lambda_j^{(m)}}t^{j-i};q)_\infty}
       {(Q_{lm}q^{\lambda_i^{(l)}-\lambda_j^{(m)}}t^{j-i+1};q)_\infty}
  \prod_{l,m}^n \prod_{i=1}^{N^{(l)}}
  \frac{(Q_{lm}q^{\lambda_i^{(l)}}t^{N^{(m)}-i+1};q)_\infty}
       {(Q_{ml}q^{-\lambda_i^{(l)}}t^{-N^{(m)}+i};q)_\infty}.
\end{equation}
Here $(x;q)_n = \prod_{m=0}^{n-1} (1-xq^m)$ is the $q$-Pochhammer
symbol, and the Coulomb moduli is denoted as $Q_{lm}=e^{a_{lm}}=q^{b_{lm}}$.
Note that this $q$-partition function includes the infinite product,
so that we have to take care of its radius of convergence.
Therefore we first consider the parametrization (\ref{qt_parametrize}),
and then take the limit $q\to 1$.
This orbifold projecting procedure is quite useful, for example, to investigate
asymptotic behavior of the orbifold partition function because it can be simply
given by studying asymptotics of the $q$-partition function in a usual way, and
taking its root of unity limit at last.

We now comment on the relation to the explicit expressions for
$\mathcal{O}_{\PP_1}(-r)$, etc, which is shown in
Refs.~\cite{Fucito:2006kn,Bruzzo:2008}.
They are written down in terms of the local coordinates of the resolved space.
For example, in the case of $\mathcal{O}_{\PP_1}(-r)$, they are given by
$(z_{1}^{(1)},z_2^{(1)})=(z_1^r, z_1^{-1}z_2)$ and
$(z_1^{(2)},z_2^{(2)}) = (z_1z_2^{-1},z_2^r)$, which are invariant under
$(z_1, z_2) \to (\omega_r z_1, \omega_r z_2)$.
Thus we have the instanton partition function, which is manifestly
invariant under the orbifold action.
This manipulation gives rise to redefinition of the partitions: if we consider
$\U(1)$ theory for simplicity, we have $\lambda_i \to \lambda_I^{(1)},
\lambda_I^{(2)}$, which satisfy $\lambda_i = r \lambda_I^{(1)} + I$ with
$i=I$ or $\lambda_i = \lambda_I^{(2)}$ with $i = \lambda_I^{(2)}+r I$.
Here the superscript labels the local patch of the resolved space.
The expressions for $\mathcal{O}_{\PP_1}(-r)$ can be obtained by these
redefined variables.
Anyway this connection is still complicated, thus it should be
investigated in detail for further study.

\begin{figure}[t]
 \begin{center}
  \includegraphics[width=20em]{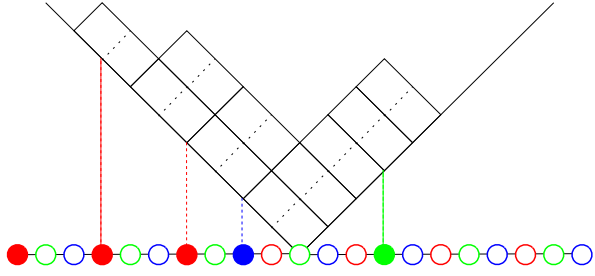}
 \end{center}
 \caption{Decomposition of the partition $\lambda=(5,2,2,1)$ for
 $\C^2/\Gamma_{3,1}$.
 A rectangular box is required for obtaining the correspondence
 between the partition and the particle description with the {\em
 repulsion} parameter $\bar{s}=r-s$.}
 \label{Young_diag}
\end{figure}

To treat the orbifold partition function more conveniently, we then try
to decompose the partitions.
In this case we introduce a slightly different way of decomposition \cite{Dijkgraaf:2007fe,Kimura:2011zf},
\begin{equation}
 r (\lambda_i^{(l,v)}+N^{(l,v)}-i+p^{(l,v)}) + v
  \equiv \lambda_j^{(l)} + \bar{s} (N^{(l)} - j) + p_l, \qquad
  j = c_i^{(l,v)}
\end{equation}
where $c_i^{(l,v)}$ stands for the mapping
from the index of the divided $nr$-partition to that of the original
$n$-partition.
Fig.~\ref{Young_diag} shows an example for the case with the orbifold
$\C^2/\Gamma_{3,1}$, namely $\bar{s}=2$.

This decomposition is based on the particle description obeying the fractional
exclusive statistics \cite{PhysRevLett.67.937}, which is deeply related
to Calogero-Sutherland model (see, for example,
Ref.~\cite{KuramotoKato200908}).
The parameter $\bar{s}$ stands for the strength of the repulsion between
particles, and this generalized statistics goes back to the usual
fermionic one in the case of the ALE space, $\bar{s}=1$.
It is useful to introduce a rectangular box to obtain the
correspondence between the partition and the particle description due to
the repulsion parameter $\bar{s}$.

Introducing another set of variables defined as
\begin{equation}
  \ell_i^{(l,v)} \equiv r (\lambda_i^{(l,v)}+N^{(l,v)}-i+p^{(l,v)}) +
   b_l - p_l + v,
\end{equation}
we finally obtain $r$-tuple partition by blending $nr$-tuple one,
\begin{equation}
 \ell^{(v)}_{i=1, \cdots, \sum_{l=1}^{n}N^{(l,v)}} = 
  \left(
   \ell_1^{(n,v)}, \cdots, \ell_{N^{(n,v)}}^{(n,v)}, \cdots,
   \ell_1^{(1,v)}, \cdots, \ell_{N^{(1,v)}}^{(1,v)}
  \right).
\end{equation}
We now assume $N^{(l)}={\bf N}$ and $N^{(l,v)}=N$ for simplicity.
Thus the partition function is rewritten in terms of $r$-tuple
partition,
\begin{eqnarray}
 Z_{\vec\lambda}^q & = & \prod_{(v,i)\not=(w,j)}
  \frac{(\omega_r^{v-w}q^{\ell_i^{(v)}-\ell_j^{(w)}
  +(\beta-\bar{s})(c_j^{(w)}-c_i^{(v)})};\omega_r q)_\infty}
  {(\omega_r^{v-w+\bar{s}}q^{\ell_i^{(v)}-\ell_j^{(w)}
  +(\beta-\bar{s})(c_j^{(w)}-c_i^{(v)})+\beta};\omega_r q)_\infty}
  \nonumber \\
 && \times \prod_{l=1}^n \prod_{v=0}^{r-1} \prod_{i=1}^{nN}
  \frac{(\omega^{r-p_l+\bar{s}}q^{\ell_i^{(v)}-b_l
  +(\beta-\bar{s})({\bf N}-c_i^{(v)})+\beta};\omega_r q)_\infty}
  {(\omega^{-r+p_l}q^{-(\ell_i^{(v)}-b_l
  +(\beta-\bar{s})({\bf N}-c_i^{(v)}))};\omega_r q)_\infty}
  \label{partfunc_SUn_orb}
\end{eqnarray}
Let $c_i^{(v)}$ stand for the mapping of the index from $r$-tuple to
$nr$-tuple partition as before.

\section{Matrix model description}\label{sec:matrix}

We then derive matrix model description by taking asymptotic limit of
the combinatorial representation of the partition function
(\ref{partfunc_SUn_orb}).
Such an integral representation would be useful to extract the gauge
theory consequences by performing the large $N$ limit analysis: the
Seiberg-Witten curve is obtained from the spectral curve of the matrix
model \cite{Dijkgraaf:2009pc,Klemm:2008yu,Kimura:2011zf}.

We now consider the following function to study the asymptotics
of the partition function,
\begin{equation}
 f_{q,t}(x) = \frac{(x;q)_\infty}{(tx;q)_\infty}.
  \label{q_func}
\end{equation}
The asymptotics of this function is almost given by the limit of $|q|\to
1$ because, when $x=e^{y}$, we have
\begin{equation}
 f_{q,t}(x) = \prod_{n=0}^\infty 
  \frac{1 - e^{y+n\epsilon_1}}{1 - e^{y+(n+\beta)\epsilon_1}}.
\end{equation}
Therefore the condition $y \gg \epsilon_1$ corresponds to the limit $q \to 1$.
This function is investigated in detail in appendix~\ref{sec:q_vander}.

Thus we now apply the result for the double root of unity limit
 (\ref{dbl_root_asymp}) to the combinatorially represented partition function.
Introducing the following variables
\begin{equation}
 x_i^{(v)} = \frac{\ell_i^{(v)}}{\epsilon_1},
\end{equation}
and taking the limit $\epsilon_1\to 0$, we then obtain the matrix model
representation which captures the asymptotics of the combinatorial
partition function.
For the orbifold $\C^2/\Gamma_{r,s}$, we have
the $r$-matrix model,
\begin{equation}
 Z = \int \mathcal{D} \vec{X}~ e^{-\frac{1}{\epsilon_1} 
  \sum_{v=0}^{r-1} \sum_{i=1}^N V(x_i^{(v)})},
\end{equation}
\begin{equation}
 \mathcal{D} \vec{X} = \prod_{v=0}^{r-1} \prod_{i=1}^N
  \frac{d x_i^{(v)}}{2\pi} \Delta^2(x).
\end{equation}
Note that we have to replace the summation over the partition with the
integral of the continuous variables. 
This is done by inserting an auxiliary function, which has a simple pole
at all integer values of the argument
\cite{Klemm:2008yu,Sulkowski:2009br,Sulkowski:2009ne}.
This affects on the matrix integral as just a linear shift of
the matrix potential in the large $N$ limit, which can be absorbed by
the counting parameter.

Let us now discuss the matrix measure and the potential function.
According to (\ref{dbl_root_asymp}), asymptotics of the first part in
(\ref{partfunc_SUn_orb}), which will go to the measure part of the
matrix model, yields
\begin{eqnarray}
 &&
 \prod_{(v,i)\not=(w,j)}
  f_{q,t}\left(
	  \omega_r^{v-w} q^{\ell_i^{(v)}-\ell_j^{(w)}
	  +(\beta-\bar{s})(c_j^{(w)}-c_i^{(v)})}
	 \right)
  \nonumber \\
  & \simeq &
  \prod_{(v,i)\not=(w,j)}
  \left[
   \left(
    1 - e^{r(x_i^{(v)}-x_j^{(w)})}
   \right)^{(\beta-\bar{s})/r}
   \prod_{k=0}^{\bar{s}-1}
   \left(
    1 - \omega_r^{v-w+k} e^{x_i^{(v)}-x_j^{(w)}}
   \right)
  \right].
\end{eqnarray}
This coincides with the following matrix measure, up to the overall factor,
\begin{eqnarray}
 \Delta^2(x) & = &
  \prod_{v=0}^{r-1} \prod_{i<j}^N
  \left[
   \left(
    2 \sinh \frac{r}{2}
    \left(x_i^{(v)} - x_j^{(v)}\right)
   \right)^{2(\beta-\bar{s})/r}
   \prod_{k=0}^{\bar{s}-1}
   \left(
    2 \sinh \frac{1}{2}
    \left(x_i^{(v)} - x_j^{(v)} + \frac{2\pi i}{r}k \right)
   \right)^{2\bar{s}}
  \right]
  \nonumber \\
 & \times &
  \prod_{v<w}^{r-1} \prod_{i, j}^N
  \left[
   \left(
    2 \sinh \frac{r}{2}
    \left(x_i^{(v)} - x_j^{(w)}\right)
   \right)^{2(\beta-\bar{s})/r}
   \prod_{k=0}^{\bar{s}-1}
      \left(
    2 \sinh \frac{1}{2}
    \left (x_i^{(v)} - x_j^{(w)} + \frac{2\pi i}{r}(v-w+k) \right)
   \right)^{2}
  \right].
  \nonumber \\
 \label{mat_meas}
\end{eqnarray}
We redefine the matrix size as $nN \to N$ for convenience.
We ramark that $\beta$ can take a generic value, thus (\ref{mat_meas})
is interpreted as the measure part of the $\beta$-ensemble matrix model
for the generic toric orbifold.

The corresponding four dimensional limit is given by
\begin{eqnarray}
 \Delta^2(x) & \to &
  \prod_{v=0}^{r-1} \prod_{i<j}^N
  \left(x_i^{(v)} - x_j^{(v)}\right)^{2(\beta-\bar{s})/r+2\bar{s}}
  \prod_{v<w}^{r-1} \prod_{i, j}^N
    \left(x_i^{(v)} - x_j^{(w)}\right)^{2(\beta-\bar{s})/r+2N_{\bar{s}}(v-w)},
    \label{vander_4d}
\end{eqnarray}
where we define
$N_{\bar{s}}(x)=\#\{k|x+k\equiv0~(\mbox{mod}~r),k=0, \cdots,
\bar{s}-1\}$.

We can easily obtain important examples from this generic result.
For the case of $\bar{s}=1$, corresponding to the orbifold $\C^2/\Z_r$, we have
\begin{eqnarray}
 \Delta^2(x) & = &
  \prod_{v=0}^{r-1} \prod_{i<j}^N
  \left[
   \left(
    2 \sinh \frac{r}{2}
    \left(x_i^{(v)} - x_j^{(v)}\right)
   \right)^{2(\beta-1)/r}
   \left(
    2 \sinh \frac{1}{2}
    \left(x_i^{(v)} - x_j^{(v)} \right)
   \right)^{2}
  \right]
  \nonumber \\
 & \times &
  \prod_{v<w}^{r-1} \prod_{i, j}^N
  \left[
   \left(
    2 \sinh \frac{r}{2}
    \left(x_i^{(v)} - x_j^{(w)}\right)
   \right)^{2(\beta-1)/r}
      \left(
    2 \sinh \frac{1}{2}
    \left (x_i^{(v)} - x_j^{(w)} + \frac{2\pi i}{r}(v-w) \right)
   \right)^{2}
  \right].
\nonumber \\  
\end{eqnarray}
This is consistent with the previous result \cite{Kimura:2011zf}.
In this time this formula is available even for generic $\beta$ while
only the specific case $\beta = r\gamma+1 \equiv 1$ (mod~$r$), $\gamma =
0, 1, 2, \cdots$, is
investigated in the previous paper.
This generalization is quite important because, in terms of AGT
relation, the parameter $\beta$ plays a crucial role in both of the four
dimensional and two dimensional theories.

The second is the case $\bar{s}=r$, corresponding to the orbifold $\C\times\C/\Z_r$,
\begin{eqnarray}
 \Delta^2(x) & = &
  \prod_{(v,i)\not=(w,j)}
  \left(
   2 \sinh \frac{r}{2}
   \left(
    x_i^{(v)} - x_j^{(w)}
   \right)
  \right)^{2\beta/r}
  \nonumber \\
 & \to &
  \prod_{(v,i)\not=(w,j)}
  \left(x_i^{(v)}-x_j^{(w)}\right)^{2\beta/r}.
\end{eqnarray}
This case is also important to study the instanton counting in presence
of the generic surface operators as discussed in Ref.~\cite{Kanno:2011fw}.

We then consider the potential part for the matrix model.
It is useful to discuss the quantum dilogarithm function
\cite{Eynard:2008mt} for deriving the matrix potential,
\begin{equation}
 g(z;q) = \prod_{p=1}^\infty \left(1-\frac{1}{z}q^p\right).
\end{equation}
In particular, when we parametrize $q=e^{\epsilon_1}$, the asymptotic
behavior at the root of unity~\cite{Kimura:2011zf} is given by
\begin{equation}
 \log g(z;\omega_r q) = 
  \frac{1}{\epsilon_1}
  \left[
   \frac{1}{r^2} {\rm Li}_2 \left(\frac{1}{z^r}\right) + \mathcal{O}(\epsilon_1)
  \right]
\end{equation}
where ${\rm Li}_2(x) = \sum_{p=1}^\infty z^p/p^2$ is the dilogarithm function.
Thus the second part in (\ref{partfunc_SUn_orb}) leads to the matrix
potential,
\begin{equation}
 \prod_{l=1}^n \prod_{v=0}^{r-1} \prod_{i=1}^{nN}
  \frac{(\omega^{r-p_l+\bar{s}}q^{\ell_i^{(v)}-b_l
  +(\beta-\s)({\bf N}-c_i^{(v)})+\beta};\omega_r q)_\infty}
  {(\omega^{-r+p_l}q^{-(\ell_i^{(v)}-b_l
  +(\beta-\s)({\bf N}-c_i^{(v)}))};\omega_r q)_\infty}
  \equiv \exp
  \sum_{v=0}^{r-1} \sum_{i=1}^{nN} - \frac{1}{\epsilon_1} V(x_i^{(v)}),
\end{equation}
\begin{equation}
 V(x) = - \frac{1}{r^2} \sum_{l=1}^n
  \left[
   {\rm Li}_2 (e^{r(x-a_l)}) - {\rm Li}_2 (e^{-r(x-a_l)})
  \right]
  + \mathcal{O}(\epsilon_1).
\end{equation}
This potential function is completely the same as the previous result
\cite{Kimura:2011zf}. 
It depends on only $r$, but $\s$ nor $\beta$.
The corresponding four dimensional limit is given by
\begin{equation}
 V(x) \longrightarrow \frac{2}{r} \sum_{l=1}^n 
  \left[
   (x-a_l) \log (x-a_l) - (x-a_l)
  \right].
\end{equation}
In this paper we concentrate on the case without the matter fields, but
it is expected that we can obtain the same matrix potential to the
homogeneous orbifolds $\C^2/\Z_r$ as well as the vector multiplet.

\section{Summary and discussion}\label{sec:summary}

In this paper we have extended the previous results \cite{Kimura:2011zf}
to the toric orbifolds $\C^/\Gamma_{r,s}$ with a generic
deformation parameter $\beta$.
The instanton counting on such an inhomogeneous orbifold would play an
essential role on the AGT relation in presence of the surface operator
\cite{Kanno:2011fw}.
Furthermore, since this parameter $\beta$ is directly related to the
$\Omega$-background parameter as $\beta=-\epsilon_2/\epsilon_1$, 
it is important to assign a generic value for the application to the
AGT relation.

We have considered the ADHM construction for the toric orbifolds
$\C^2/\Gamma_{r,s}$, and derived the instanton partition
function for such a space.
We have shown that the root of unity limit is
useful to implement the orbifold projection.
It has been also shown that the partition function is well described by
the particles obeying the fractional exclusive statistics for the
generic case.

Based on such a combinatorial description, we have obtained the
corresponding $\beta$-ensemble multi-matrix models by considering its
asymptotic behavior.
The matrix measure is directly related to the root of unity limit of the
$q$-deformed Vandermonde determinant, and reflecting the structure of
the orbifolds $\C^2/\Gamma_{r,s}$.
On the other hand, the matrix potential depends on only $r$, but $s$.

We concentrate on obtaining the matrix model description in this paper,
but we do not deal with the matrix model itself in detail.
Actually the matrix model, which we have derived, has an apparently
complicated expression.
However, this matrix model is obtained by the non-standard reduction
of the $q$-deformed theory, which should be integrable because it can be
represented in terms of the $q$-free boson fields.
Thus it is expected that there is an integrable structure even for our
matrix model.
In the large $N$ limit we could perform the standard treatment of the
matrix model as well as the generic lens space matrix model
\cite{Brini:2008ik}, and obtain the corresponding Seiberg-Witten curve
as the spectral curve.
Furthermore, the relation between the model discussed in this paper and
another kind of matrix model, i.e. Dijkgraaf-Vafa's model
\cite{Dijkgraaf:2009pc}, is worth studying in detail, because the latter
plays an essential role in the AGT relation.
It is one of the possibilities of further study beyond this work.

It is also interesting to discuss the corresponding two dimensional
conformal field theory to the generic toric orbifold theory.
Since the inhomogeneous orbifold theory is utilized to study the instanton
partition function in presence of a surface operator
\cite{Kanno:2011fw}, it is expected to obtain a similar structure for
the generic toric orbifold theory, corresponding to the
para-Liouville/Toda theory \cite{Nishioka:2011jk}.
It would provide a novel perspective to explore an exotic conformal field
theory.

\section*{Acknowledgments}

The author would like to thank T.~Nishioka and Y.~Tachikawa for valuable
comments.
The author is supported by Grant-in-Aid for JSPS Fellows.

\appendix
\section{Reduction of the $q$-Vandermonde determinant}\label{sec:q_vander}

The function (\ref{q_func}) is directly related to the weight function
of the Macdonald polynomial \cite{Mac_book}, namely the $q$-deformed
Vandermonde determinant, 
\begin{equation}
 \Delta_{q,t}^2(x) = \prod_{i\not=j} 
  \frac{(x_i/x_j;q)_\infty}{(tx_i/x_j;q)_\infty}
  = \prod_{i\not=j} f_{q,t}(x_i/x_j).
\end{equation}
According to the $q$-binomial theorem we have
\begin{equation}
 f_{q,t}^{-1}(x)
  = \sum_{n=0}^\infty \frac{(t;q)_n}{(q;q)_n} x^n.
  \label{q-exp1}
\end{equation}
In this appendix we investigate several kinds of reduction of the
$q$-deformed Vandermonde determinant.

The first example is given by the following parametrization,
\begin{equation}
 t = q^\beta, \qquad
  q \longrightarrow 1.
\end{equation}
Because the coefficient becomes
\begin{equation}
 \frac{(t;q)_n}{(q;q)_n} \longrightarrow (-1)^n 
  \left(
   \begin{array}{c}
    -\beta \\ n
   \end{array}
  \right),
\end{equation}
we have
\begin{equation}
 f_{q,t}^{-1}(x) \longrightarrow 
  \sum_{n=0}^\infty (-x)^n
  \left(
   \begin{array}{c}
    -\beta \\ n
   \end{array}
  \right)
  = (1-x)^{-\beta}.
\end{equation}
This corresponds to the Jack limit of the Macdonald polynomial since 
the $q$-Vandermonde is reduced to
\begin{equation}
 \Delta_{q,t}^2(x) \longrightarrow
  \prod_{i\not=j}
  \left(1-\frac{x_i}{x_j}\right)^{\beta}
  \sim \prod_{i<j} (x_i-x_j)^{2\beta}.
\end{equation}
The same kind of reduction is found for the $q$-Virasoro algebra
\cite{Shiraishi:1995rp}, which leads to the usual Virasoro algebra with
the central charge $c=1-6(\beta-1)^2/\beta$.

The next is the single root of unity limit, which is used to study the
instanton counting on the orbifold $\C^2/\Z_r$ \cite{Kimura:2011zf}, and
corresponds to the parametrization proposed in \cite{Uglov:1997ia},
\begin{equation}
 q \longrightarrow \omega_r q, \qquad
 t \longrightarrow \omega_r q^\beta, \qquad
 q \longrightarrow 1.
\end{equation}
The expansion coefficient in (\ref{q-exp1}) is given by
\begin{eqnarray}
 \frac{(t;q)_n}{(q;q)_n} & = &
  \prod_{m=1}^n 
  \frac{1-\omega_r^m t q^{m-1}}{1-\omega_r^m q^m}
  \nonumber \\
 & \longrightarrow & 
  \prod_{m=1}^{[n/r]}
  \frac{\beta+rm-1}{rm}
  \nonumber \\
 & = & (-1)^{\left[n/k\right]}
  \left(
   \begin{array}{c}
    - \left(\frac{\beta-1}{r}+1\right) \\ \left[n/r\right]
   \end{array}
  \right).
\end{eqnarray}
Here $[x]$ denotes the largest integer not greater than $x$.
Therefore we have
\begin{eqnarray}
 f_{q,t}^{-1}(x) & \longrightarrow &
  \sum_{n=0}^\infty (-x^r)^n 
  \left(
   \begin{array}{c}
    - \left(\frac{\beta-1}{r}+1\right) \\ n
   \end{array}
  \right)
  \left(
   1 + x + \cdots + x^{r-1}
  \right)
  \nonumber \\
 & = & 
  (1-x)^{-1} \left(1-x^r\right)^{-(\beta-1)/r}.
\end{eqnarray}
As a result, the $q$-Vandermonde is reduced as
\begin{equation}
 \Delta_{q,t}^2(x) \longrightarrow
  \prod_{i\not=j} 
  \left(1-\frac{x_i}{x_j}\right) 
  \left(1-\frac{x_i^r}{x_j^r}\right)^{(\beta-1)/r}
  \sim
  \prod_{i<j}
  (x_i - x_j)^2
  \left( x_i^r - x_j^r \right)^{2(\beta-1)/r}.
\end{equation}
This is consistent with the previous result \cite{Uglov:1997ia,Kimura:2011zf}.
Note that this reduction is available for generic positive $\beta$ while
only the specific case $\beta=r\gamma+1\equiv 1$ (mod~$r$) has been
investigated so far.

The last is the double root of unity limit.
We now consider the following parametrization,
\begin{equation}
 q \longrightarrow \omega_r q, \qquad
 t \longrightarrow \omega_r^{\s} q^\beta, \qquad
 q \longrightarrow 1.
\end{equation}
The coefficient in
(\ref{q-exp1}) becomes
\begin{eqnarray}
 \frac{(t;q)_n}{(q;q)_n} & \longrightarrow &
  \prod_{m=1}^{\s-1} \frac{1-\omega_r^{n+m}}{1-\omega_r^m}
  \prod_{m=1}^{[n/r]} \frac{(\beta-\s)/r+m}{m}
  \nonumber \\
 & = &
  (-1)^{[n/r]}
  \left(
   \begin{array}{c}
    -\left(\frac{\beta-\s}{r}+1\right) \\ \left[n/r\right]
   \end{array}
  \right)
  \prod_{m=1}^{\s-1} \frac{1-\omega_r^{n+m}}{1-\omega_r^m} .
\end{eqnarray}
Note that this coefficient vanishes as $(t;q)_n/(q;q)_n = 0$ when $n
\equiv r-\s+1, \cdots, r-1$ (mod~$r$).
Thus we have a similar result,
\begin{equation}
  f_{q,t}^{-1}(x) \longrightarrow 
   (1-x^r)^{-(\beta-\s)/r} \prod_{k=0}^{\s-1} (1-\omega_r^k x)^{-1}.
   \label{dbl_root_asymp}
\end{equation}
This corresponds to the following reduction of the $q$-Vandermonde,
\begin{eqnarray}
 \Delta_{q,t}^2(x) & \longrightarrow &
  \prod_{i\not=j} 
  \left(1-\frac{x_i^r}{x_j^r}\right)^{(\beta-\s)/r}
  \prod_{k=0}^{\s-1} \left(1-\omega_r^k \frac{x_i}{x_j}\right).
\end{eqnarray}
Especially, when $\s=r$, corresponding to the orbifold $\C\times\C/\Z_r$,
which is well investigated in \cite{Kanno:2011fw}, it becomes
\begin{equation}
 \Delta_{q,t}^2(x) \longrightarrow
  \prod_{i\not=j} \left(1-\frac{x_i^r}{x_j^r}\right)^{\beta/r}
  \sim
  \prod_{i<j} (x_i^r - x_j^r)^{2\beta/r}.
\end{equation}


\bibliographystyle{ytphys}

\bibliography{/Users/k_taro/Configure/conf}

\providecommand{\href}[2]{#2}\begingroup\raggedright\begin{thebibliography}{10}

\bibitem{Nekrasov:2002qd}
N.~Nekrasov, ``{Seiberg-Witten Prepotential From Instanton Counting},'' {\em
  Adv. Theor. Math. Phys.} {\bfseries 7} (2004) 831--864,
\href{http://arxiv.org/abs/hep-th/0206161}{{\ttfamily arXiv:hep-th/0206161}}.

\bibitem{Nekrasov:2003rj}
N.~Nekrasov and A.~Okounkov, ``{Seiberg-Witten theory and random partitions},''
\href{http://arxiv.org/abs/hep-th/0306238}{{\ttfamily arXiv:hep-th/0306238}}.

\bibitem{Seiberg:1994rs}
N.~Seiberg and E.~Witten, ``{Monopole condensation, and confinement in ${\cal
  N}=2$ supersymmetric Yang-Mills theory},''
  \href{http://dx.doi.org/10.1016/0550-3213(94)90124-4}{{\em Nucl. Phys.}
  {\bfseries B426} (1994) 19--52},
\href{http://arxiv.org/abs/hep-th/9407087}{{\ttfamily arXiv:hep-th/9407087}}.

\bibitem{Seiberg:1994aj}
N.~Seiberg and E.~Witten, ``{Monopoles, duality and chiral symmetry breaking in
  ${\cal N}=2$ supersymmetric QCD},''
  \href{http://dx.doi.org/10.1016/0550-3213(94)90214-3}{{\em Nucl. Phys.}
  {\bfseries B431} (1994) 484--550},
\href{http://arxiv.org/abs/hep-th/9408099}{{\ttfamily arXiv:hep-th/9408099}}.

\bibitem{Alday:2009aq}
L.~F. Alday, D.~Gaiotto, and Y.~Tachikawa, ``{Liouville Correlation Functions
  from Four-dimensional Gauge Theories},''
  \href{http://dx.doi.org/10.1007/s11005-010-0369-5}{{\em Lett. Math. Phys.}
  {\bfseries 91} (2010) 167--197},
\href{http://arxiv.org/abs/0906.3219}{{\ttfamily arXiv:0906.3219 [hep-th]}}.

\bibitem{Wyllard:2009hg}
N.~Wyllard, ``{$A_{N-1}$ conformal Toda field theory correlation functions from
  conformal $\mathcal{N}=2$ $SU(N)$ quiver gauge theories},''
  \href{http://dx.doi.org/10.1088/1126-6708/2009/11/002}{{\em JHEP} {\bfseries
  11} (2009) 002},
\href{http://arxiv.org/abs/0907.2189}{{\ttfamily arXiv:0907.2189 [hep-th]}}.

\bibitem{Mironov:2009by}
A.~Mironov and A.~Morozov, ``{On AGT relation in the case of U(3)},''
  \href{http://dx.doi.org/10.1016/j.nuclphysb.2009.09.011}{{\em Nucl. Phys.}
  {\bfseries B825} (2010) 1--37},
\href{http://arxiv.org/abs/0908.2569}{{\ttfamily arXiv:0908.2569 [hep-th]}}.

\bibitem{Gaiotto:2009ma}
D.~Gaiotto, ``{Asymptotically free $\mathcal{N}=2$ theories and irregular
  conformal blocks},''
\href{http://arxiv.org/abs/0908.0307}{{\ttfamily arXiv:0908.0307 [hep-th]}}.

\bibitem{Marshakov:2009gn}
A.~Marshakov, A.~Mironov, and A.~Morozov, ``{On non-conformal limit of the AGT
  relations},'' \href{http://dx.doi.org/10.1016/j.physletb.2009.10.077}{{\em
  Phys. Lett.} {\bfseries B682} (2009) 125--129},
\href{http://arxiv.org/abs/0909.2052}{{\ttfamily arXiv:0909.2052 [hep-th]}}.

\bibitem{Taki:2009zd}
M.~Taki, ``{On AGT Conjecture for Pure Super Yang-Mills and W-algebra},''
  \href{http://dx.doi.org/10.1007/JHEP05(2011)038}{{\em JHEP} {\bfseries 05}
  (2011) 038},
\href{http://arxiv.org/abs/0912.4789}{{\ttfamily arXiv:0912.4789 [hep-th]}}.

\bibitem{Belavin:2011pp}
V.~Belavin and B.~Feigin, ``{Super Liouville conformal blocks from
  $\mathcal{N}=2$ SU(2) quiver gauge theories},''
  \href{http://dx.doi.org/10.1007/JHEP07(2011)079}{{\em JHEP} {\bfseries 07}
  (2011) 079},
\href{http://arxiv.org/abs/1105.5800}{{\ttfamily arXiv:1105.5800 [hep-th]}}.

\bibitem{Nishioka:2011jk}
T.~Nishioka and Y.~Tachikawa, ``{Para-Liouville/Toda central charges from
  M5-branes},'' \href{http://dx.doi.org/10.1103/PhysRevD.84.046009}{{\em Phy.
  Rev.} {\bfseries D84} (2011) 046009},
\href{http://arxiv.org/abs/1106.1172}{{\ttfamily arXiv:1106.1172 [hep-th]}}.

\bibitem{Bonelli:2011jx}
G.~Bonelli, K.~Maruyoshi, and A.~Tanzini, ``{Instantons on ALE spaces and Super
  Liouville Conformal Field Theories},''
  \href{http://dx.doi.org/10.1007/JHEP08(2011)056}{{\em JHEP} {\bfseries 08}
  (2011) 056},
\href{http://arxiv.org/abs/1106.2505}{{\ttfamily arXiv:1106.2505 [hep-th]}}.

\bibitem{Belavin:2011tb}
A.~Belavin, V.~Belavin, and M.~Bershtein, ``{Instantons and 2d Superconformal
  field theory},'' \href{http://dx.doi.org/10.1007/JHEP09(2011)117}{{\em JHEP}
  {\bfseries 09} (2011) 117},
\href{http://arxiv.org/abs/1106.4001}{{\ttfamily arXiv:1106.4001 [hep-th]}}.

\bibitem{Bonelli:2011kv}
G.~Bonelli, K.~Maruyoshi, and A.~Tanzini, ``{Gauge Theories on ALE Space and
  Super Liouville Correlation Functions},''
\href{http://arxiv.org/abs/1107.4609}{{\ttfamily arXiv:1107.4609 [hep-th]}}.

\bibitem{Eguchi:1978xp}
T.~Eguchi and A.~J. Hanson, ``{Asymptotically Flat Selfdual Solutions to
  Euclidean Gravity},''
  \href{http://dx.doi.org/10.1016/0370-2693(78)90566-X}{{\em Phys. Lett.}
  {\bfseries B74} (1978) 249}.

\bibitem{Gibbons:1979zt}
G.~Gibbons and S.~Hawking, ``{Gravitational Multi-Instantons},''
  \href{http://dx.doi.org/10.1016/0370-2693(78)90478-1}{{\em Phys. Lett.}
  {\bfseries B78} (1978) 430}.

\bibitem{Kronheimer:1989zs}
P.~B. Kronheimer, ``{The Construction of ALE Spaces as hyper-K\"ahler
  Quotients},''
\href{http://projecteuclid.org/euclid.jdg/1214443066}{{\em J. Diff. Geom.}
  {\bfseries 29} (1989) 665--683}.

\bibitem{springerlink:10.1007/BF01233429}
H.~Nakajima, ``{Moduli spaces of anti-self-dual connections on ALE
  gravitational instantons},'' \href{http://dx.doi.org/10.1007/BF01233429}{{\em
  Invent. Math.} {\bfseries 102} (1990) 267--303}.

\bibitem{springerlink:10.1007/BF01444534}
P.~B. Kronheimer and H.~Nakajima, ``{Yang-Mills instantons on ALE gravitational
  instantons},'' \href{http://dx.doi.org/10.1007/BF01444534}{{\em Math. Ann.}
  {\bfseries 288} (1990) 263--307}.

\bibitem{Fucito:2004ry}
F.~Fucito, J.~F. Morales, and R.~Poghossian, ``{Multi instanton calculus on ALE
  spaces},'' \href{http://dx.doi.org/10.1016/j.nuclphysb.2004.09.014}{{\em
  Nucl. Phys.} {\bfseries B703} (2004) 518--536},
\href{http://arxiv.org/abs/hep-th/0406243}{{\ttfamily arXiv:hep-th/0406243}}.

\bibitem{Nishinaka:2011nn}
T.~Nishinaka and S.~Yamaguchi, ``{Affine $SU(N)$ algebra from
  wall-crossings},'' \href{http://arxiv.org/abs/1107.4762}{{\ttfamily
  arXiv:1107.4762 [hep-th]}}.

\bibitem{Nishinaka:2011is}
T.~Nishinaka and Y.~Yoshida, ``{A note on statistical model for BPS D4-D2-D0
  states},''
\href{http://arxiv.org/abs/1108.4326}{{\ttfamily arXiv:1108.4326 [hep-th]}}.

\bibitem{Kanno:2011fw}
H.~Kanno and Y.~Tachikawa, ``{Instanton counting with a surface operator and
  the chain-saw quiver},''
  \href{http://dx.doi.org/10.1007/JHEP06(2011)119}{{\em JHEP} {\bfseries 06}
  (2011) 119},
\href{http://arxiv.org/abs/1105.0357}{{\ttfamily arXiv:1105.0357 [hep-th]}}.

\bibitem{Kimura:2011wh}
T.~Kimura and M.~Nitta, ``{Vortices on Orbifolds},''
  \href{http://dx.doi.org/10.1007/JHEP09(2011)118}{{\em JHEP} {\bfseries 09}
  (2011) 118},
\href{http://arxiv.org/abs/1108.3563}{{\ttfamily arXiv:1108.3563 [hep-th]}}.

\bibitem{Kimura:2011zf}
T.~Kimura, ``{Matrix model from $\mathcal{N} = 2$ orbifold partition
  function},'' \href{http://dx.doi.org/10.1007/JHEP09(2011)015}{{\em JHEP}
  {\bfseries 09} (2011) 015},
\href{http://arxiv.org/abs/1105.6091}{{\ttfamily arXiv:1105.6091 [hep-th]}}.

\bibitem{Fucito:2006kn}
F.~Fucito, J.~F. Morales, and R.~Poghossian, ``{Instanton on toric
  singularities and black hole countings},''
  \href{http://dx.doi.org/10.1088/1126-6708/2006/12/073}{{\em JHEP} {\bfseries
  12} (2006) 073},
\href{http://arxiv.org/abs/hep-th/0610154}{{\ttfamily arXiv:hep-th/0610154}}.

\bibitem{Griguolo:2006kp}
L.~Griguolo, D.~Seminara, R.~J. Szabo, and A.~Tanzini, ``{Black holes,
  instanton counting on toric singularities and q-deformed two-dimensional
  Yang-Mills theory},''
  \href{http://dx.doi.org/10.1016/j.nuclphysb.2007.02.030}{{\em Nucl. Phys.}
  {\bfseries B772} (2007) 1--24},
\href{http://arxiv.org/abs/hep-th/0610155}{{\ttfamily arXiv:hep-th/0610155}}.

\bibitem{Brini:2008ik}
A.~Brini, L.~Griguolo, D.~Seminara, and A.~Tanzini, ``{Chern-Simons theory on
  $L(p,q)$ lens spaces and Gopakumar-Vafa duality},''
  \href{http://dx.doi.org/10.1016/j.geomphys.2009.11.006}{{\em J. Geom. Phys.}
  {\bfseries 60} (2010) 417--429},
\href{http://arxiv.org/abs/0809.1610}{{\ttfamily arXiv:0809.1610 [math-ph]}}.

\bibitem{Gang:2009wy}
D.~Gang, ``{Chern-Simons theory on $L(p,q)$ lens spaces and Localization},''
\href{http://arxiv.org/abs/0912.4664}{{\ttfamily arXiv:0912.4664 [hep-th]}}.

\bibitem{Benini:2011nc}
F.~Benini, T.~Nishioka, and M.~Yamazaki, ``{4d Index to 3d Index and 2d
  TQFT},''
\href{http://arxiv.org/abs/1109.0283}{{\ttfamily arXiv:1109.0283 [hep-th]}}.

\bibitem{Uglov:1997ia}
D.~Uglov, ``{Yangian Gelfand-Zetlin bases, $\mathfrak{gl}_N$-Jack polynomials
  and computation of dynamical correlation functions in the spin
  Calogero-Sutherland model},''
  \href{http://dx.doi.org/10.1007/s002200050283}{{\em Commun. Math. Phys.}
  {\bfseries 193} (1998) 663--696},
\href{http://arxiv.org/abs/hep-th/9702020}{{\ttfamily arXiv:hep-th/9702020}}.

\bibitem{KuramotoKato200908}
Y.~Kuramoto and Y.~Kato, \href{http://dx.doi.org/10.1017/CBO9780511596827}{{\em
  {Dynamics of One-Dimensional Quantum Systems: Inverse-Square Interaction
  Models}}}.
\newblock Cambridge University Press, 2009.

\bibitem{Klemm:2008yu}
A.~Klemm and P.~Su{\l}kowski, ``{Seiberg-Witten theory and matrix models},''
  \href{http://dx.doi.org/10.1016/j.nuclphysb.2009.04.004}{{\em Nucl. Phys.}
  {\bfseries B819} (2009) 400--430},
\href{http://arxiv.org/abs/0810.4944}{{\ttfamily arXiv:0810.4944 [hep-th]}}.

\bibitem{Sulkowski:2009br}
P.~Su{\l}kowski, ``{Matrix models for $2^*$ theories},''
  \href{http://dx.doi.org/10.1103/PhysRevD.80.086006}{{\em Phys. Rev.}
  {\bfseries D80} (2009) 086006},
\href{http://arxiv.org/abs/0904.3064}{{\ttfamily arXiv:0904.3064 [hep-th]}}.

\bibitem{Sulkowski:2009ne}
P.~Su{\l}kowski, ``{Matrix models for $\beta$-ensembles from Nekrasov partition
  functions},'' \href{http://dx.doi.org/10.1007/JHEP04(2010)063}{{\em JHEP}
  {\bfseries 04} (2010) 063},
\href{http://arxiv.org/abs/0912.5476}{{\ttfamily arXiv:0912.5476 [hep-th]}}.

\bibitem{Mac_book}
I.~G. Macdonald, {\em {Symmetric Functions and Hall Polynomials}}.
\newblock Oxford University Press, 2nd~ed., 1997.

\bibitem{BarthPetersVandeven}
W.~Barth, C.~Peters, and A.~Van~de Ven, {\em {Compact Complex Surfaces}}.
\newblock Springer-Verlag, 1984.

\bibitem{Nakajima:2003uh}
H.~Nakajima and K.~Yoshioka, ``{Lectures on instanton counting},''
\href{http://arxiv.org/abs/math/0311058}{{\ttfamily arXiv:math/0311058}}.

\bibitem{Gasparim:2008ri}
E.~Gasparim and C.-C.~M. Liu, ``{The Nekrasov Conjecture for Toric Surfaces},''
  \href{http://dx.doi.org/10.1007/s00220-009-0948-4}{{\em Commun. Math. Phys.}
  {\bfseries 293} (2010) 661--700},
\href{http://arxiv.org/abs/0808.0884}{{\ttfamily arXiv:0808.0884 [math.AG]}}.

\bibitem{Bruzzo:2008}
U.~Bruzzo, R.~Poghossian, and A.~Tanzini, ``{Instanton counting on Hirzebruch
  surfaces},'' \href{http://arxiv.org/abs/0809.0155}{{\ttfamily arXiv:0809.0155
  [math.AG]}}.

\bibitem{Dijkgraaf:2007fe}
R.~Dijkgraaf and P.~Su{\l}kowski, ``{Instantons on ALE spaces and orbifold
  partitions},'' \href{http://dx.doi.org/10.1088/1126-6708/2008/03/013}{{\em
  JHEP} {\bfseries 03} (2008) 013},
\href{http://arxiv.org/abs/0712.1427}{{\ttfamily arXiv:0712.1427 [hep-th]}}.

\bibitem{PhysRevLett.67.937}
F.~D.~M. Haldane, ``{``Fractional statistics'' in arbitrary dimensions: A
  generalization of the Pauli principle},''
  \href{http://dx.doi.org/10.1103/PhysRevLett.67.937}{{\em Phys. Rev. Lett.}
  {\bfseries 67} (1991) 937--940}.

\bibitem{Dijkgraaf:2009pc}
R.~Dijkgraaf and C.~Vafa, ``{Toda Theories, Matrix Models, Topological Strings,
  and $N=2$ Gauge Systems},''
\href{http://arxiv.org/abs/0909.2453}{{\ttfamily arXiv:0909.2453 [hep-th]}}.

\bibitem{Eynard:2008mt}
B.~Eynard, ``{All orders asymptotic expansion of large partitions},''
  \href{http://dx.doi.org/10.1088/1742-5468/2008/07/P07023}{{\em J. Stat.
  Mech.} {\bfseries 07} (2008) P07023},
\href{http://arxiv.org/abs/0804.0381}{{\ttfamily arXiv:0804.0381 [math-ph]}}.

\bibitem{Shiraishi:1995rp}
J.~Shiraishi, H.~Kubo, H.~Awata, and S.~Odake, ``{A Quantum deformation of the
  Virasoro algebra and the Macdonald symmetric functions},''
  \href{http://dx.doi.org/10.1007/BF00398297}{{\em Lett. Math. Phys.}
  {\bfseries 38} (1996) 33--51},
\href{http://arxiv.org/abs/q-alg/9507034}{{\ttfamily arXiv:q-alg/9507034}}.

\end{thebibliography}\endgroup

\end{document}